\documentclass[%
 aip,
 apl,
 amsmath,amssymb,
 reprint,%
]{revtex4-1}

\usepackage{CJK}
\usepackage{stackengine}
\usepackage{graphicx}
\usepackage{subfigure}
\usepackage{color}
\usepackage{lipsum}
\usepackage{dcolumn}
\usepackage{bm}
\usepackage{textcomp}
\usepackage[euler]{textgreek}
\usepackage{ctable}

\begin{document}
\begin{CJK}{UTF8}{}
\preprint{AIP/123-QED}

\title[Sample title]{Eliminating the non-Gaussian spectral response of X-ray absorbers for transition-edge sensors}

\author{Daikang Yan (\CJKfamily{gbsn}{闫代康})}

\affiliation{Argonne National Laboratory, Argonne, Illinois 60439, USA}
\affiliation{Northwestern Univeristy, Evanston, Illinois 60208, USA}

\author{Ralu Divan}
\author{Lisa M. Gades}
\author{Peter Kenesei}
\author{Timothy J. Madden}
\author{Antonino Miceli}

\author{Jun-Sang Park}
\author{Umeshkumar M. Patel}
\affiliation{Argonne National Laboratory, Argonne, Illinois 60439, USA}

\author{Orlando Quaranta}
\affiliation{Argonne National Laboratory, Argonne, Illinois 60439, USA}
\affiliation{Northwestern Univeristy, Evanston, Illinois 60208, USA}

\author{Hemant Sharma}
\affiliation{Argonne National Laboratory, Argonne, Illinois 60439, USA}

\author{Douglas A. Bennett}
\author{William B. Doriese}
\author{Joseph W. Fowler}
\affiliation{National Institute of Standards and Technology, Boulder, Colorado 80305, USA}

\author{Johnathon Gard}
\affiliation{National Institute of Standards and Technology, Boulder, Colorado 80305, USA}
\affiliation{University of Colorado, Boulder, Colorado 80309, USA}

\author{James Hays-Wehle}
\affiliation{National Institute of Standards and Technology, Boulder, Colorado 80305, USA}

\author{Kelsey M. Morgan}
\affiliation{National Institute of Standards and Technology, Boulder, Colorado 80305, USA}
\affiliation{University of Colorado, Boulder, Colorado 80309, USA}

\author{Daniel R. Schmidt}
\author{Daniel S. Swetz}
\affiliation{National Institute of Standards and Technology, Boulder, Colorado 80305, USA}

\author{Joel N. Ullom}
\affiliation{National Institute of Standards and Technology, Boulder, Colorado 80305, USA}
\affiliation{University of Colorado, Boulder, Colorado 80309, USA}

\date{\today}

\begin{abstract}
Transition-edge sensors (TES) as microcalorimeters for high-energy-resolution X-ray spectroscopy are often fabricated with an absorber made of materials with high \textit{Z} (for X-ray stopping power) and low heat capacity (for high resolving power). Bismuth represents one of the most compelling options. TESs with evaporated bismuth absorbers have shown spectra with undesirable and unexplained low-energy tails. We have developed TESs with electroplated bismuth absorbers over a gold layer that are not afflicted by this problem and that retain the other positive aspects of this material. To better understand these phenomena, we have studied a series of TESs with gold, gold/evaporated bismuth, and gold/electroplated bismuth absorbers, fabricated on the same die with identical thermal coupling. We show that bismuth morphology is linked to the spectral response of X-ray TES microcalorimeters.
\end{abstract}

\maketitle
\end{CJK}

X-ray transition-edge sensor (TES) microcalorimeters are superconducting devices that measure X-ray photon energy using the sharp resistance variation upon photon-induced temperature change. It has long been demonstrated that X-ray TES microcalorimeters have eV-scale energy resolution and have nearly reached their theoretical limits \cite{Ullom:2005bl}. However, in addition to energy resolution, an important performance metric for X-ray TESs is the expected Gaussian shape of the spectral response. The most common non-Gaussian response observed is a low-energy (LE) tail associated with each X-ray emission line. This nonideality complicates X-ray line-shape analysis and degrades the detectability of trace element mapping \cite{Sun:2015fr}. Because the efficient detection of X-rays generally requires thick absorbers, high-atomic-number and low-heat-capacity materials are desirable. Bismuth (Bi) is one such material for X-ray TES absorbers that is widely used. In particular, Bi is a compelling material compared to gold (Au) because it has comparable X-ray detection quantum efficiency but an order of magnitude smaller specific heat capacity, arising from the low carrier density of Bi \cite{PhysRevB.1.2888}. Both evaporated \cite{Doriese:2017ex} and electroplated \cite{Bandler:2008fv} Bi have been extensively used for X-ray TESs. In addition, bulk, single-crystal Bi glued to silicon microcalorimeters \cite{Stahle:1993vu} has also been examined. While the non-Gaussian response of X-ray TESs has been observed predominately in evaporated Bi absorbers \cite{Tatsuno:2016el}, the response of an X-ray TES depends on not only the properties of the absorber itself, but also the thermal coupling between the TES and absorber. In this work, we disentangle these two issues. In particular, we present a systematic study of three types of absorbers with identical thermal coupling to the TES.

\begin{figure}[!]
	\centering
	\begin{minipage}[!]{0.45\textwidth}
	   \topinset{\bfseries{\textcolor{white}{(a)}}}{
       \includegraphics[width=1\textwidth]{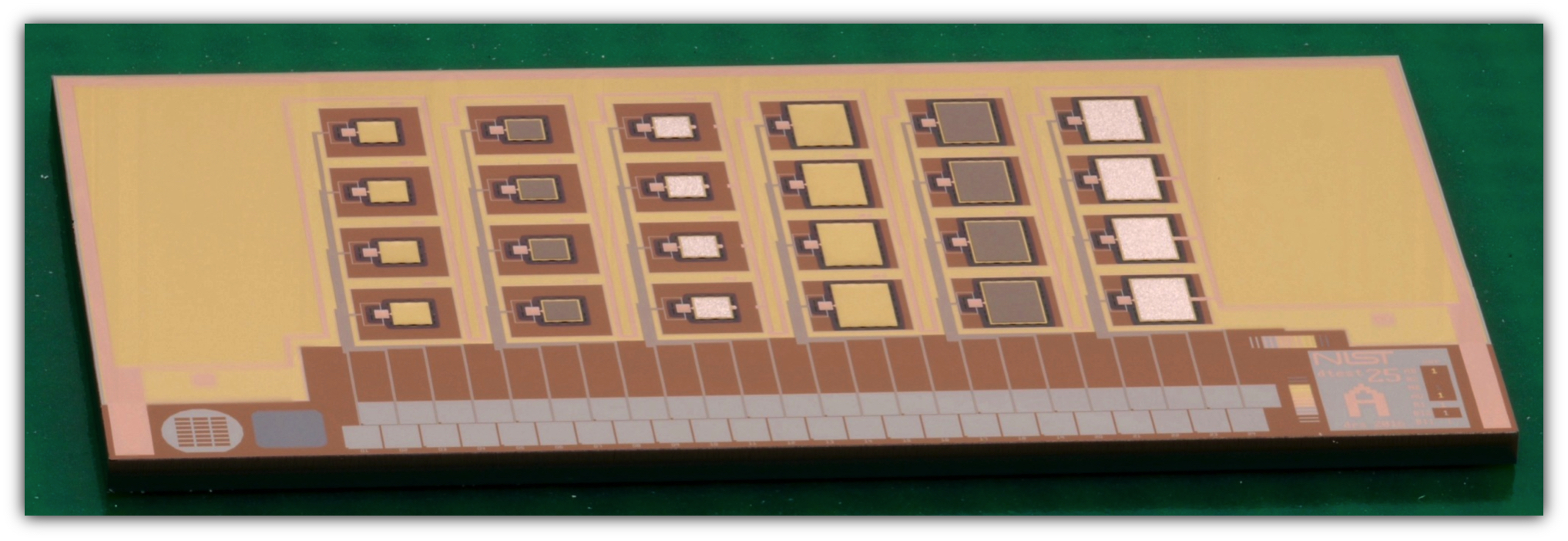}  
       \label{fig:TES_photo} 
        }{0.05\textwidth}{-0.4\textwidth}
  \end{minipage}
	\vfill
	\begin{minipage}[!]{0.45\textwidth}
	   \topinset{\bfseries(b)}{
       \includegraphics[width=1\textwidth]{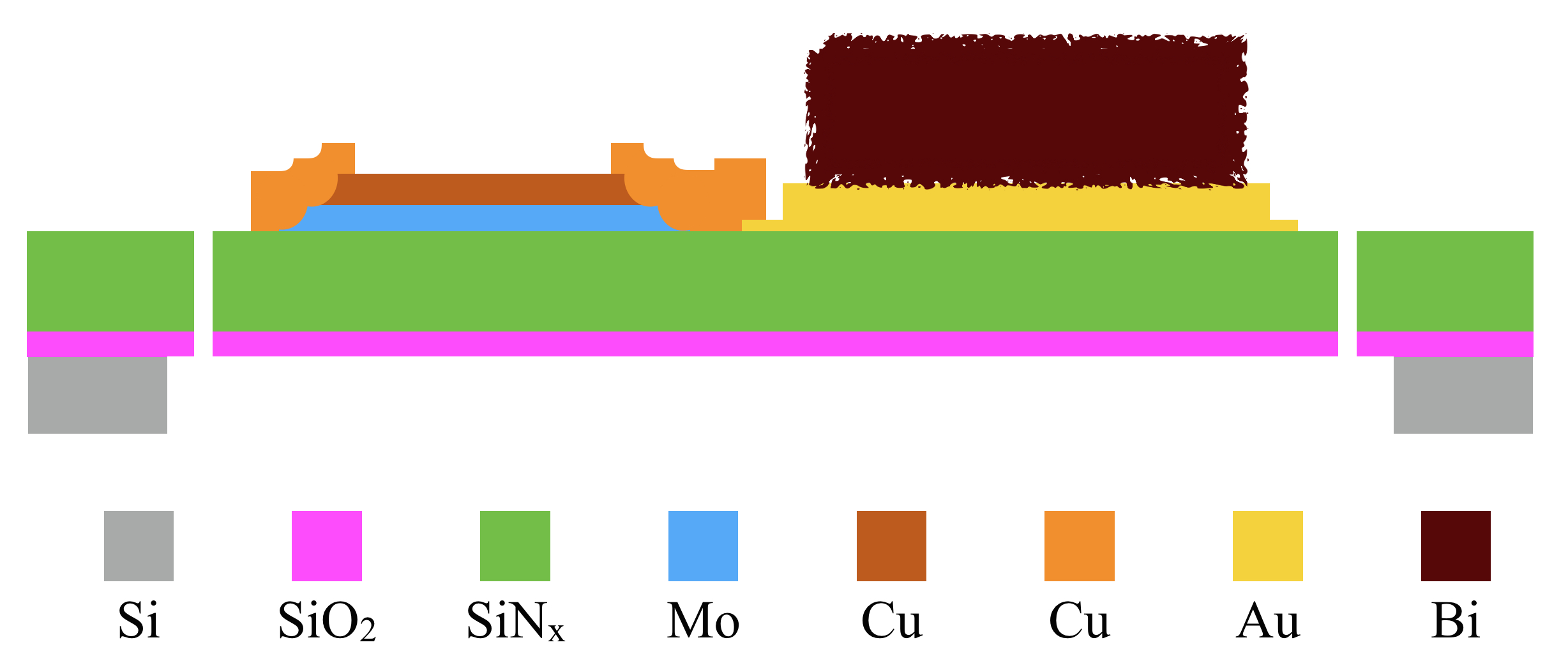}  
       \label{fig:pixel_side}
        }{0.05\textwidth}{-0.4\textwidth}
  \end{minipage}
	\caption{(a): Photograph of the TES die (7 mm x 14 mm). There are two sets of devices with small and large area absorbers designed for X-rays ranging up to $\sim$ 10 keV and $\sim$ 30 keV, respectively. For each device size there are three types of absorbers, with differing materials, from left to right: Au, Au/evap-Bi and Au/elp-Bi. For all absorbers the Au is 1 \textmu m thick, and the Bi is 3 \textmu m thick. (b): Schematic representation of the cross-sectional view of a device (not to scale).}
	\label{fig:device_picture}
\end{figure}

We have designed a TES that is compatible with electroplated (elp) and evaporated (evap) depositions (Fig.~\ref{fig:device_picture}). The TES is a molybdenum/copper (Mo/Cu) bilayer with Cu banks and bars on top for transition-parameter control and noise suppression \cite{Ullom:2005bl}. 
The transition temperature of these TESs is $\sim$ 100 mK. The absorbers are positioned to the side of the TES and attached to the Cu banks via a 0.2 \textmu m thick Au layer. An additional 0.8 \textmu m of Au forms the base of the absorber. 
The evap-Bi is deposited via a lift-off process, while the elp-Bi is electrodeposited \cite{Gades:2017ix} via a patterned Au seed layer with a Cu current path that is subsequently removed. 
Each type of Bi absorber is 3 \textmu m thick, which offers $\sim$ 76\% quantum efficiency at 6 keV and $\sim$ 8\% at 30 keV. The devices are grouped in two sets, designed for small (340 x 340 \textmu m\textsuperscript{2}) and large (530 x 720 \textmu m\textsuperscript{2}) absorbers, with dynamic range up to $\sim$ 10 keV and $\sim$ 30 keV, respectively. 
For each set, the thermal conductance (\textit{G}) is optimized to maintain electrothermal stability of the TES while biased. \textit{G} is controlled via a perforated SiN\textsubscript{x} membrane that supports the entire device (Fig.~\ref{fig:device_picture}(b)). 
In each set, there are three types of devices, with differing absorbers, from left to right: Au, Au/evap-Bi and Au/elp-Bi. Finally, there are four copies of each. 

\begin{table*}[!]
\begin{tabular}{c c c c c c c}
\specialrule{0.05em}{0.1em}{0.1em} 
\specialrule{0.1em}{0em}{0em} 
{Size}&\multicolumn{3}{c}{Small pixel}&\multicolumn{3}{c}{Large pixel}\\ 
{Pixel perimeter} (\textmu m) & \multicolumn{3}{c}{1860} & \multicolumn{3}{c}{3000}\\ 
\specialrule{0.05em}{0.1em}{0.1em} 
Absorber material & Au & Au/Evap-Bi & Au/Elp-Bi & Au & Au/Evap-Bi & Au/Elp-Bi\\
\specialrule{0em}{0.1em}{0.1em} 
\textit{G} (pW/K) & 254.7 & 254.3 & 263.0 & 384.0 & 392.3 & 400.3\\
\specialrule{0em}{0.1em}{0.1em} 
\textit{C} (pJ/K) & 1.2 & 1.2 & 1.1 & 3.1 & 2.9 & 3.0\\
\specialrule{0.1em}{0.1em}{0em}
\specialrule{0.05em}{0.1em}{0.1em} 
\end{tabular}
\caption{Representative thermal conductances and heat capacities of the different type of pixels.}
\label{tab:tab_param}
\end{table*}

The \textit{G} from the pixel to the heat bath was measured from the Joule-heating power at different bath temperatures with the TES biased at 85\% of the normal state resistance. 
As expected, \textit{G} scaled with the perimeter of the TES plus the absorber (Tab.~\ref{tab:tab_param}), as is expected in the case where the energy transport is dominated by specular reflections of phonons at the SiN\textsubscript{x} surfaces. 
Next, the devices were characterized using an X-ray generator exciting fluorescence from several metallic foils. An aperture confined the X-ray illumination area of each device to the absorber. 
The device's total heat capacity (\textit{C}) was calculated from the relationship \textit{C} = \textit{G} $\times$ $\bm{\tau}$\textsubscript{thermal}, where $\bm{\tau}$\textsubscript{thermal} was approximated by the pulse-decay time $\bm{\tau}$ at \textit{T}\textsubscript{b} $\approx$ \textit{T}\textsubscript{c}, where the TES resistance dependence on temperature and current is small. 
The measured \textit{C} values scaled with the absorber volumes only and did not depend on the type of Bi used. Moreover, the contribution from the Bi was negligible, as shown in Tab.~\ref{tab:tab_param}. 

\begin{figure}[t]
    \centering
    \begin{minipage}[!]{0.23\textwidth}
        \centering 
        \topinset{\bfseries(a)}{\includegraphics[width=\textwidth]{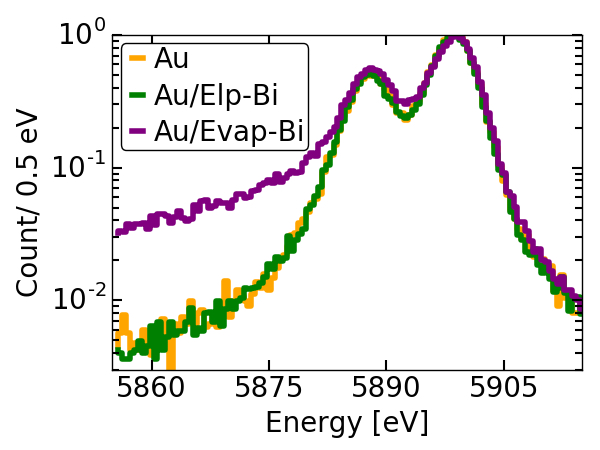}}{0.1\textwidth}{0.37\textwidth}
        \label{fig:hist_compare}
    \end{minipage}
    \vspace{-1em}    
    \begin{minipage}[!]{0.23\textwidth}  
        \centering 
        \topinset{\bfseries(b)}{\includegraphics[width=\textwidth]{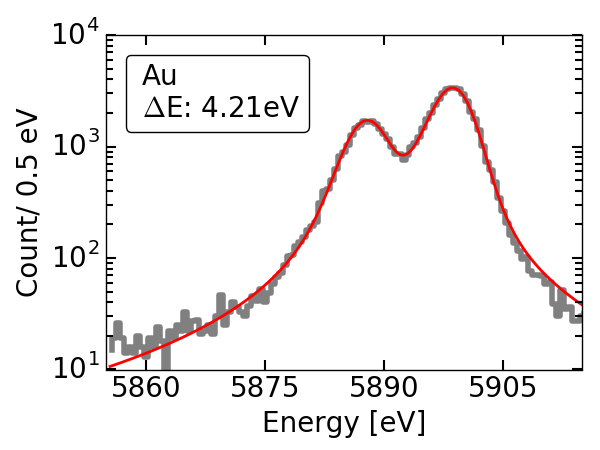}}{0.1\textwidth}{0.37\textwidth}
        \label{fig:hist_Au}
    \end{minipage}
    \vspace{-1em}    
    \begin{minipage}[!]{0.23\textwidth}   
        \centering 
        \topinset{\bfseries(c)}{\includegraphics[width=\textwidth]{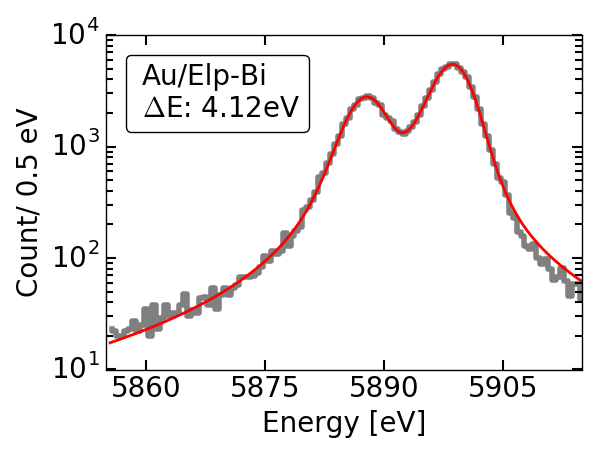}}{0.1\textwidth}{0.37\textwidth}
        \label{fig:hist_elpBi}
    \end{minipage}
    \begin{minipage}[!]{0.23\textwidth}   
        \centering 
        \topinset{\bfseries(d)}{\includegraphics[width=\textwidth]{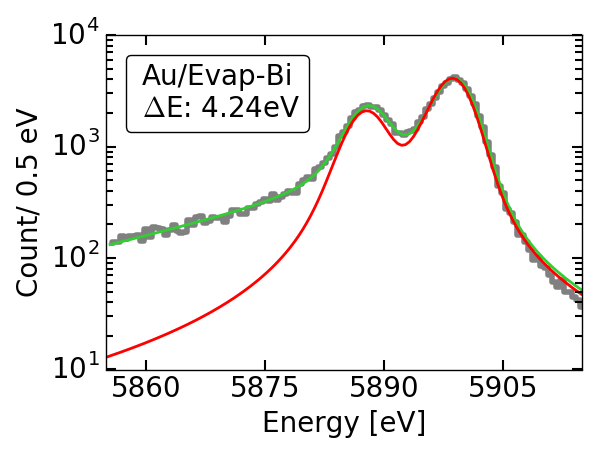}}{0.1\textwidth}{0.37\textwidth}
        \label{fig:hist_evapBi}
    \end{minipage}
    \caption{(a): Comparison of the Mn {K\textalpha} spectrum measured by small pixels with the three types of absorbers (normalized at the peak maxima). The Au/evap-Bi absorber devices show a LE tail compared with the Au and Au/elp-Bi absorbers. (b) and (c): Measured spectra for Au and Au/elp-Bi respectively (black) are well-matched to a Gaussian fit (red). (d): Measured spectrum from Au/evap-Bi (black) is poorly matched by a simple Gaussian fit (red), while a Gaussian with tail fit (green) better approximates it.} 
    \label{fig:hist}
\end{figure}

\begin{figure}[b]
\centering
\includegraphics[width=0.45\textwidth]{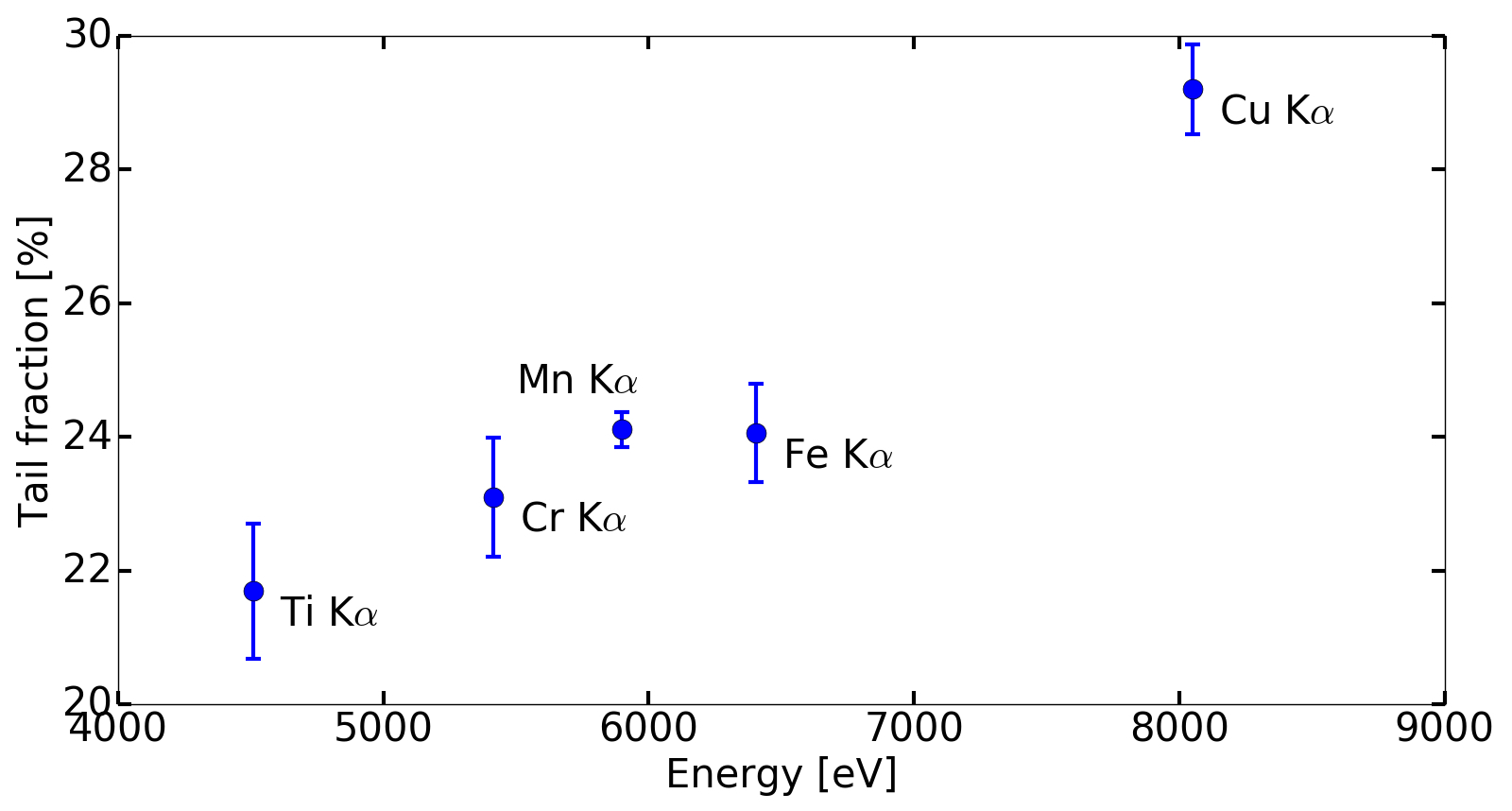}
\caption{The LE-tail fraction of the Ti, Cr, Mn, Fe and Cu {K\textalpha} lines for a small Au/evap-Bi device. Error bars denote the fit standard deviation.}
\label{fig:tail}
\end{figure}

This was to be expected considering that the specific heat capacity of Bi is one order of magnitude smaller than that of Au. Because the energy resolution is proportional to $\sqrt{C}$, the Bi layer is not expected to introduce any energy resolution penalty.
X-ray pulse heights (i.e., energies) were estimated via optimal-filter-based techniques \cite{Fowler:2015is}. The Mn {K\textalpha} spectra in Fig.~\ref{fig:hist}(a) measured by small Au and Au/elp-Bi pixels are nearly identical, while the Au/evap-Bi spectrum shows a clear LE tail. 
The spectra from the three types of pixels show similar energy resolution, which is consistent with the heat capacity measurement. 
To further examine if a subtle LE tail exists in the Au and Au/elp-Bi spectra, we fitted our observed spectra to a convolution of a Gaussian detector response function and the natural line shape, as previously measured by H\"olzer, et al. \cite{Holzer:1997bn}, represented by the red lines in Figs.~\ref{fig:hist}(b)-(d). 
Both the Au and Au/elp-Bi data match well with the Gaussian response function, indicating no presence of LE tails. Conversely, the simple Gaussian fit fails when applied to the Au/evap-Bi spectrum. 
The LE tail in the Au/evap-Bi spectrum can be fitted by the additional convolution of an exponential function with a Gaussian distribution (green line in Fig.~\ref{fig:hist}(d)), as suggested in Ref.~\onlinecite{Tatsuno:2016el}. In Fig.~\ref{fig:tail}, measurements of the LE tail fraction, which denotes the portion of energy deposited in the tail, from a small Au/evap-Bi absorber for titanium (Ti), chromium (Cr), manganese (Mn), iron (Fe) and Cu {K\textalpha} emissions are reported (a similar trend is observed in the large absorber). A clear increase in the LE tail fraction with X-ray energy is present; this has also been reported by Tatsuno, et al. \cite{Tatsuno:2016el} and Fowler, et al. \cite{Fowler:2017cw}.

The LE tail is indicative of some mechanism that prevents measurement of the full energy deposited in the absorber, with consequent generation of a current pulse that is smaller than expected. 
The missing energy either escapes from or is trapped within the evap-Bi and not transformed into thermal energy over a time scale that is short compared to the TES typical response time ($\sim$ 1 ms). 
Typical energy-escape mechanisms are either emission of a fluorescence photon by the absorber, which does not get re-absorbed by it, or escape of phonons or electrons, generated in the energy relaxation process, through the substrate or the surface to the ambient environment. 
From the analysis of the LE-tail characteristic and its dependence on the incident photon energy, we can exclude the hypothesis of escape photons. 
The fluorescence photons emitted by Bi are at discrete energies ({M\textalpha} at 2.423 keV and {M\textbeta} at 2.525 keV) \cite{Fowler:2017cw}, which is incompatible with the continuous distribution of the LE tail seen in the evap-Bi spectrum. 
Moreover, the escape fluorescence mechanism would be inherent to Bi, regardless of the deposition technique. The loss of energy through the substrate is also unlikely because 1 \textmu m of Au is sufficient to thermalize energetic phonons \cite{Kozorezov:2011kn} from either type of absorber. In addition, the use of the SiN\textsubscript{x} membrane greatly inhibits the escape of high-energy phonons; those that reach the membrane are likely to reflect off its surface and be reabsorbed by the Au or Au/Bi. Furthermore, Kilbourne, et al. \cite{Kilbourne} report that both devices with an evaporated Bi absorber in direct contact with the SiN\textsubscript{x} membrane and with the Bi in contact only with the TES show the presence of the LE tail. 
The electron photoemission is also unlikely to happen because the photon-absorption event occurs deep in the absorber - the average attenutation length for Bi at these energies is several \textmu m. Moreover, the higher the energy of the incident photon the deeper the absorption is likely to happen, making the electron escape harder, which contradicts the increase of LE tail fraction with energy in Fig.~\ref{fig:tail}. Finally, the work functions for Au and Bi are comparable (Au $\sim$ 5 eV, Bi $\sim$ 4 eV), which would make the probability of a photoemission event quite similar in the two materials at these X-ray photon energies. Based on these considerations, energy-escape mechanisms are improbable.

\begin{figure}[b]
\centering
	\begin{minipage}[!]{0.23\textwidth}
	   \topinset{\bfseries{\textcolor{white}{(a)}}}{\includegraphics[width=1\textwidth]{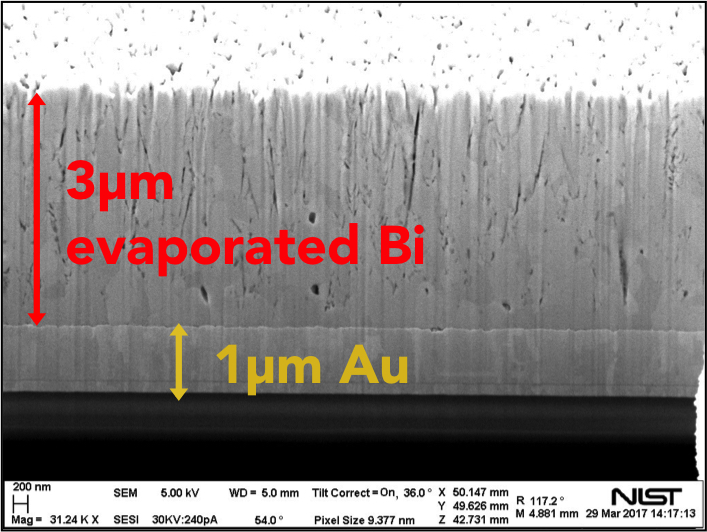}}{0.58\textwidth}{0.4\textwidth}
	   \label{fig:evapBi_SEM} 
	\end{minipage}
	\hfill
	\begin{minipage}[!]{0.23\textwidth}
	   \topinset{\bfseries{\textcolor{white}{(b)}}}{\includegraphics[width=1\textwidth]{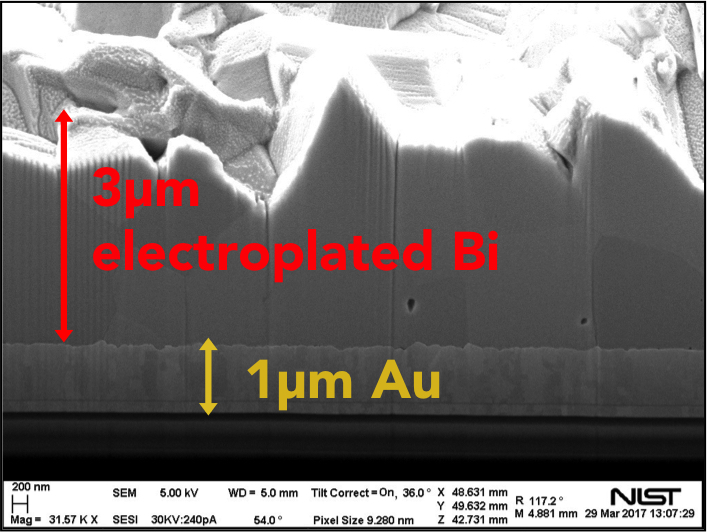}}{0.58\textwidth}{0.4\textwidth}
	   \label{fig:elpBi_SEM}
	\end{minipage}
	\caption{Scanning electron micrographs of the (a): evap-Bi and (b): elp-Bi absorber cross-sections. The evap-Bi grain appears smaller than that of the elp-Bi, and shows a columnar structure.}
	\label{fig:SEM}
\end{figure}

In semimetal Bi, the dominant contributions to heat transport from different heat carriers change with temperature, and non-lattice contributions to the total thermal conductivity can be significant at very low temperatures ($\sim$ 100 mK) relevant to TES operation \cite{Uher:1985wc}. 
Fig.\ref{fig:SEM}(a) and Fig.\ref{fig:SEM}(b) show focused-ion-beam cross-sectional images of the Au/evap-Bi and Au/elp-Bi absorbers obtained via a scanning electron microscope. The evap-Bi grains show a columnar structure, and appear to be much smaller than the elp-Bi grains. 
The grain sizes were measured using high energy X-ray diffraction at the 1-ID-E beamline of the Advanced Photon Source, Argonne National Laboratory \cite{Lienert:2011et}. 
The evap-Bi resulted in a powder pattern, and its full width at half maxima fitted from the diffraction peaks were used in the Scherrer equation to estimate the average grain size \cite{Patterson:1939gp}, which was approximately 30 nm. The elp-Bi resulted in distinct diffraction spots, indicating larger Bi grains. By taking the ratio between the volume of absorber illuminated by the X-ray and the total number of diffraction spots normalized for multiplicity \cite{Sharma:2012gt}, the average grain radius in the elp-Bi absorber was approximately 630 nm.

Our data on transport measurements for evap-Bi show semiconductor-like behavior (i.e., residual resistance ratio of 0.4) due to grain boundary scattering \cite{Yang}, in agreement with the measured grain size. 
This further reduces the non-lattice thermal conductivity due to scattering of heat carriers at the grain boundaries and crystal defects; charge carrier thermal conductivity scales inversely with resistivity. The larger number of grain boundaries in the evap-Bi is likely to scatter or trap heat carriers, and the columnar structure may constrain the heat from transferring horizontally, causing difficulty in the thermalization of the entire absorber. Moreover, Doriese, et al.  \cite{Doriese:2017ex} report that the LE tail fraction of the spectrum increases with thickness of the evaporated film. This supports the idea that the thermalization mechanism in these columnar films happens mainly in the vertical direction, therefore the further an event is from the underlying Au layer, the more probable is the heat trapping mechanism. 
The Bi semimetal-to-semiconductor transition occurs when the energy shift due to quantum confinement raises the lowest electron sub-band to an energy higher than the uppermost hole sub-band \cite{Hoffman1993, Lin:2000ug}. 
This transition happens for sizes below 50 nm, and the energy gap has been shown to depend on the grain size \cite{Dresselhaus:1998bg, Lin:2000bg}.  
If the nature of the grains in evap-Bi is semiconductive, this could also trap the heat in the form of long-lived electrons excited above the semiconducting energy gap. 
Given the grain size of the evap-Bi, we can relate the LE tail to this size-induced phase transition effect.
It is also conceivable that bismuth oxide may be present at the grain boundaries. As an insulator, bismuth oxide could result in long-lived electron-hole pairs that behave like lost energy.  

The dependence of the LE tail fraction on the incident photon energy seen in Fig.\ref{fig:tail} and other work \cite{Doriese:2017ex, Tatsuno:2016el} supports the hypothesis of a morphology-based energy trapping mechanism in the evap-Bi. In particular, the size of the secondary-electron (SE) cloud generated by the incident photon in the Bi absorber is $\sim$ 60 nm at 6 keV \cite{GrumGrzhimailo:2017hv,Tabata:1996tt},  which is comparable with the grain size of the evap-Bi. Moreover, the size of the SE cloud in the absorber is proportional to the energy of the incident photon \cite{GrumGrzhimailo:2017hv}. Consequently, the number of grains (and grain boundaries) affected by the SE cloud depends on the energy of the incident photon, therefore influencing the number of events characterized by an incomplete energy collection. In contrast, the nature of our elp-Bi absorbers at these thicknesses is more similar to a classical metal with residual resistance ratio of the order of one \cite{Gades:2017ix} and thus immune from the LE tail phenomena. 

In summary, we have characterized the X-ray spectra produced by TES absorbers composed of Au, Au/evap-Bi and Au/elp-Bi at different energies. The elp-Bi showed no sign of the undesired non-Gaussian response typical of the evap-Bi and, at the same time, the amount of Bi used caused no energy-resolution degradation. The tail-free feature is especially useful to detect weak X-ray emission lines in samples with a complex matrix of materials, making the elp-Bi a superior material for large arrays of X-ray TES microcalorimeters. 
Future work to grow thicker elp-Bi absorbers with near-unity quantum efficiency up to 20 keV will increase the sensitivity of hard X-ray TESs for chemical state microscopy at the nanoscale \cite{Holt:2013dj}.

\bigbreak

This work was supported by the Accelerator and Detector R\&D program in Basic Energy Sciences' Scientific User Facilities (SUF) Division at the Department of Energy. This research used resources of the Advanced Photon Source and Center for Nanoscale Materials, U.S. Department of Energy Office of Science User Facilities operated for the DOE Office of Science by the Argonne National Laboratory under Contract No. DE-AC02-06CH11357. The contribution of NIST is not subject to copyright.

\nocite{*}
\bibliography{Bi_LEtail}

\end{document}